\documentclass[useAMS]{mn2e}

\usepackage{txfonts,amssymb}
\usepackage{amsfonts}
\usepackage{epsfig}

\newlength{\colwidth}
\def\msun{{\rm M_{\odot}}}

\title [Evolution of planets in transition discs]
{Evolutionary constraints on the planetary hypothesis for   
transition discs}
\author[C.J. Clarke \& J.E. Owen]
{C.J. Clarke$^{1}$ and J.E. Owen $^{2}$\\
$^1$Institute of Astronomy, Madingley Rd, Cambridge, CB3 0HA, UK \\ 
$^2$Canadian Institute for Theoretical Astrophysics, 60 St. George Street, Toronto,
M5S 3H8,Canada}

\pagerange{\pageref{firstpage}--\pageref{lastpage}} \pubyear{2003}

\begin{document}
\def\lta{\mathrel{\spose{\lower 3pt\hbox{$\mathchar"218$}}
     \raise 2.0pt\hbox{$\mathchar"13C$}}}
\def\gta{\mathrel{\spose{\lower 3pt\hbox{$\mathchar"218$}}
     \raise 2.0pt\hbox{$\mathchar"13E$}}}
\def\Msun{{\rm M}_\odot}
\def\msun{{\rm M}_\odot}
\def\Rsun{{\rm R}_\odot}
\def\Lsun{{\rm L}_\odot}
\def\19{GRS~1915+105}
\label{firstpage}
\maketitle

\begin{abstract}
 We assume a scenario in which  transition discs (i.e.
discs around young stars that have signatures of cool dust but lack
significant near infra-red emission from warm dust) are associated
with the presence of  
planets (or brown dwarfs). These are assumed to filter the dust content 
of any gas flow within   
the planetary orbit and produce an inner `opacity hole'. In order
to match the properties of transition discs with the largest
($\sim 50$ A.U. scale) holes, we place such  
`planets' at large radii in massive  discs 
and then follow the evolution of the tidally coupled
disc-planet system,  comparing the system's evolution 
in the plane of mm flux against hole radius 
with the  properties of observed  transition discs.
We find that, on account of the high disc masses in these systems, 
all but the most massive `planets' ($100$ Jupiter masses) are
conveyed to small radii by Type II migration without significant
fading at millimetre wavelengths. Such behaviour would contradict
the observed lack of mm bright transition discs with small
($<10$ A.U.) holes. On the other hand, imaging surveys clearly rule
out the presence of such massive companions in transition discs.
We conclude that this is a serious problem for models that seek
to explain transition discs in terms of planetary companions
unless some mechanism can be found to halt inward migration
and/or suppress mm flux production. We suggest that the dynamical effects of  
substantial accretion on to the planet/through the gap may offer
the best prospect for halting such migration but that further
long term simulations are required to clarify this issue.
\end{abstract}


\section{Introduction}

 In recent years, space based infrared observations have permitted the
identification of a large sample of {\it transition discs} (e.g. 
Najita et al 2007,Cieza et al 2008,2010; Espaillat et al 2010, Kim et al 2009, Merin et al 2010, Muzzerolle et al 2010), young
stars with spectral evidence for cool circumstellar dust but which lack 
diagnostics of warm dust. The standard interpretation
is that transition discs contain a (at least partially) cleared
inner cavity and that the temperature at the cavity wall sets 
the wavelength beyond which a strong spectral excess is detected.

The current census of transition discs totals
around $100$, which is large enough to permit some examination of trends and correlations
within the sample (e.g Alexander \& Armitage 2009, Kim et al 2009, Owen et al 2011,2012,  Owen \& Clarke 2012). A number of authors have noted that cooler
transition discs (i.e. those with a spectral upturn long-ward of $24 \mu$m,
corresponding to cavity walls at around $100$ K) are associated with 
systematically higher accretion rates on to the star and systematically
higher millimetre fluxes than warmer (small hole) transition discs.  Owen \& Clarke
(2012) demonstrated that if one divides the transition disc sample at
a mm flux that is equal to the median for young disc bearing
stars then the mm bright sub-sample has distinctly different
properties i.e. larger hole radii ($> 20$ A.U.) and higher
accretion rates compared with
 the mm faint
sample (which is dominated by small holes and a wide range of accretion rates).
Given that there appears to be
no correlation between these properties and the mm flux within
each of the two sub-samples, Owen \& Clarke (2012) suggested that these
may represent two distinct populations of transition discs. They
also noted that the properties of the mm faint sub-sample
were consistent with 
discs that were being cleared at late times as a result of Xray photoevaporation.

  The properties of the mm bright sub-sample are clearly
incompatible with disc clearing (e.g. by photoevaporation) 
at a late evolutionary stage.
Indeed the high mm fluxes imply disc masses that are in several cases
$\sim 10\%$ of the stellar mass (Andrews et al 2011); this places them in the 
regime 
that is believed to correspond to the early stages
of disc evolution and such discs may even be self-gravitating.  

 One possibility for creating a transition disc
at this stage is via the 
formation of a giant planet/brown dwarf at large radius,  which
then 
tidally truncates the disc just beyond the orbit of the planet (Kraus \& 
Ireland 2012, Nayakshin 2013). 
Although the time-scale for planet formation by core accretion is long
in the outer disc, this is a region where massive discs
can undergo gravitational fragmentation on account of their low ratio of cooling
to dynamical time-scale 
(Rafikov 2005, Stamatellos \& Whitworth 2008,
Clarke 2009).
In order to reproduce the observed high accretion rates in these
systems it is necessary that some gas can leak inwards past the planet.
On the
other hand, in order to produce  the spectral signature of a transition
disc,  this leakage flow has to be  depleted in dust.
Rice et al (2006),  Pinilla et al (2012) have suggested that such
`transparent accretion' can be effected via trapping of
dust grains at the inner edge of the tidally truncated disc.
Although 
many details of the mechanism are still to be quantified, an orbiting
companion 
provides a qualitatively attractive scenario for explaining these
objects.

  In this Letter we provisionally assume that the mm bright
transition discs are indeed created by embedded planets. We then investigate
how such systems (i.e. planet plus outer disc plus leaky
accretion flow) would evolve over the subsequent
lifetime of the disc. For this simple experiment we neglect the possible
role of photoevaporation; see Rosotti et al (2013) for a modeling of combined
photoevaporation and planet formation.  
We also assume that the level of
the leaky accretion
flow from the outer disc is around $10 \%$ of the viscous accretion
rate in the outer disc, though  - as we discuss in Section 2  below - this
value, and its influence  on planet migration, is not currently well calibrated numerically.
We furthermore assume that the dust filtration mechanism works for
all companion masses and at all orbital radii.

  We do not attempt a detailed population synthesis of transition disc
properties, due
to the large number of model assumptions  and degeneracies  in fitting
the data. Moreover, we do not {\it require} that evolution of the mm bright
transition discs (whose properties provide the initial conditions
for our experiment) can necessarily account for all  mm faint 
transition disc objects, since some of these may well have a quite different
origin (e.g. photoevaporation). What we {\it do} require is that
the model evolution does not  populate 
`forbidden' regions  of parameter space. Specifically
 we need to
a) avoid the production of large numbers of mm bright sources
with small (warm) holes and b) avoid the production of mm faint
 large holes among  conventional transition
discs (i.e. those without stellar mass companions). We add this 
proviso concerning binary companions since discs in 
wide {\it stellar} binaries have typical mm fluxes that 
are at least an order of magnitude fainter than those of
transition discs without binary  companions (Kraus et al 2011,2012; see
Figure 11, Andrews et al 2011). We therefore require that our
model generates large mm faint holes only in the limit of high
companion masses.  
\section {A simple model for coupled disc/planet evolution}

  We model the evolution of the disc according to the viscous
diffusion equation:

\begin{equation} 
{{\partial \Sigma}\over{\partial t}} = {{1}\over{R}} {{\partial}\over{\partial R}} \biggl[ 3 R^{1/2} {{\partial}\over{\partial R}} (\nu_d \Sigma R^{1/2})\biggr]
\end{equation}

\noindent where $\Sigma$ is the disc surface density and $\nu_d$ is the
kinematic viscosity which we model phenomenologically as a power law
of radius following Lynden-Bell \& Pringle 1974,
Hartmann et al 1998; here we adopt $\nu_d \propto
R$ (noting that this implies that in steady state the disc surface
density profile scales as $R^{-1}$ - cf Hartmann et al. 1998; Andrews et al  2009).
We model the coupled evolution of disc and planet through
the disc's inner boundary condition (see below); a free outflow  
condition is imposed
at the disc's outer edge.  The equation is integrated using a standard
explicit finite difference method, equispaced in $R^{1/2}$; we typically
employ $1000$ radial gridpoints over the range $1-3200$ A.U.. We have experimented
with values of the outer boundary in order to ensure that the disc
mass leaving the outer boundary is a small fraction of the initial
disc mass (a few per cent or less) and that the evolution is independent
of the outer boundary location in this case.

 We first describe the set-up in the absence of leakage from the outer disc.
 If the planet is located at grid point $i$, the inner
edge of the disc is located at grid point $i+1$, where we impose
a zero mass flux boundary condition. We then record the increase in angular
momentum of the disc resulting from  this boundary condition 
until the
total angular momentum acquired
by the disc is equal to the difference in angular momentum
of the planet in Keplerian orbit at grid points $i$ and $i-1$. (Note
that in recording the increase of angular momentum of the disc
we also include the angular momentum that is advected through
the outer boundary, where a zero torque boundary condition is
applied). At this
point, the planet is moved to grid point $i-1$, the inner edge of the
disc to grid point $i$ and the process repeated. This simple approach
ensures that the angular momentum of the system is conserved to machine
accuracy and does not - as in  the approach more usually adopted - rely on
a parametrisation of the torque between the disc and planet. We do not
expect our method to 
model  the
detailed structure of the disc in the region where it is tidally
sculpted by the planet and this will have some effect on the mm emission
(although probably not greater in magnitude than the effect of varying
the dust to gas ratio in this region, which is an effect that we do explore).
 Since  we are  interested in the orbital evolution of the planet
and  the global evolution of the disc (inasfar as this affects the mm
flux), our  simple angular momentum conserving approach
is 
sufficient for our purposes.

   In addition, we implement a leakage flow from
the outer disc. Other phenomenological modelling exercises 
(e.g. Alexander \& Armitage 2009, Alexander \& Pascucci 2012) have used
a prescription in which 
the leakage flow rises with decreasing planet mass to attain a  maximum 
of around a third of the mass flow
rate through the outer disc for planets of around  a Jupiter mass.
The appropriate values are however rather uncertain based on the existing
simulation data (Veras \& Armitage 2004,
Lubow \& D'Angelo 2006): as discussed below, in cases where the leakage
flow (and accretion on to the planet) is significant, 
there is considerable uncertainty about the
consequences for planet orbital migration inasmuch as this would deviate
from the Type II planetary migration induced purely by interaction with the
outer disc. In order to avoid these uncertainties (and because the
focus of our investigation will end up being in the more massive
planetary regime where leakage is expected to be fairly minor) 
we simply assume that the leakage flow is around $10 \%$ of the
flow in the outer disc for all companion masses.
We  cannot rule out that the
leakage flow might not become much more significant in the case of planets
of much lower planet mass and return to this issue in Section 5.

  The leakage has three consequences for the system: a) it implies
a finite accretion rate on to the star,
b) it modifies the disc evolution by depleting the outer disc and c) it
affects the planetary migration, both via b) and via the torques imparted
to the planet from the planetary accretion stream and the  flow to
the inner disc. Note that  the efficiency factor of the  leakage flow
($\epsilon$; i.e. the ratio of the leakage flow to the accretion
rate in the outer disc) )  critically determines the accretion
rate onto the star 
for all values of $\epsilon$; leakage is also significant in reducing
the millimetre flux from the disc (b)).
However  
c) is only mildly affected by leakage for low values of $\epsilon$ 
such as the value of $\epsilon ( = 0.1)$ adopted here.
This is fortunate given the uncertainties in c). 
Calculation of c) involves knowledge of the change in specific angular
momentum of fluid elements that are either directly accreted onto the
planet or are able to cross the planetary orbit into the inner disc. In
addition, the finite angular momentum possessed by material in the latter
category is eventually passed back to the planet via tidal torques at the
outer edge of the inner disc. Since  we are not modelling either the 
inner disc nor the detailed trajectories of the material crossing the
planet's orbit nor accretion onto the planet, we simply assume that the 
the entire angular momentum of the
material leaving he inner edge of the outer disc is added to 
the planet. For $\epsilon = 0.1$, relaxation of this assumption makes
negligible difference to the orbital migration of the planet which
is set almost entirely by the transfer of angular momentum to the outer
disc. 
 
  We use the instantaneous properties of the disc to compute the
mm flux, adopting standard opacity values:

\begin{equation}
\kappa_{\nu} = 0.1 \biggl({{\nu}\over{10^{12} \rm Hz}}\biggr) \rm cm^2 g^{-1}  
\end{equation}

\noindent  such that $\kappa_\nu = 0.02$ cm$^2$ g$^{-1}$ at $1.3$mm and compute the
luminosity  density (for a face-on disc) as:

\begin{equation}
L_\nu = 4 \pi {\int_{R{\rm in}}^{R_{\rm out}}\!\!\!\!\!{\rm d}R\, 2 \pi R\,  B_\nu(T(R)) (1 - e^{-\tau_\nu(R)}) } 
\end{equation}

\noindent (Beckwith et al 1990) where $B_\nu$ is the Planck function and $\tau_\nu (= \kappa_\nu \Sigma$)
is the optical depth.  We adopt a simple power law parametrisation
of the disc temperature:

\begin{equation}
T(R) = 100 \rm K \,\biggl({{R}\over{1 \rm A.U.}}\biggr)^{-0.5}
\end{equation}

\noindent which is motivated by typical parameters that have been found to
provide a fit to the spectral energy distributions of circumstellar
discs (Andrews \& Williams 2005,2007; Andrews et al. 2011; Beckwith et al 1990). 

  We explore four model discs, in all case adjusting the normalisation
of the surface density profile in order that the initial disc
has a $1.3$ mm flux (scaled to the
distance of Taurus, i.e. $140$ pc) of around $100$ mJy; thus in each
case the initial disc has properties that are typical of the mm
bright transition discs with large holes. (We emphasise that
throughout we only consider the mm flux from material that
is still in the outer disc, assuming  that dust filtration
suppresses the mm emission from the leakage flow.) None of the
results presented here depend on the normalisation of the viscosity
(since this determines the time-scale of evolution rather than the
relationship between millimetre flux and hole size that we
explore here). It is however worth noting that if we normalise
the viscosity such that the initial accretion rate {\it onto the star}
is $\sim 10^{-8}
M_\odot$ yr$^{-1}$, as observed, then the time-scale on which
the hole size shrinks to $10$ A.U. is a few Myr. We also note that
the models do not involve the mass of the star except inasmuch as this
would, in practice, affect the temperature normalisation of the disc
profile (which we have taken directly from observations; Andrews et al 2009). 

\begin{table}
\centering
\caption{Initial model parameters (see text for details)}
\begin{tabular}{rlll}
\hline
\hline
${\rm Model}$ & $R_{in} (A.U.)$ & $R_{out} (A.U.)$ & $M_{disc}$ \\
\hline
E&$50$&$150$&$50 M_{\rm{Jup}}$ \\
N&$50$&$75$&$40 M_{\rm{Jup}}$ \\
P1&$25$&$75$& $80 M_{\rm{Jup}}$  \\
P2&$25$&$75$&$30 M_{\rm{Jup}}$ \\
\hline
\end{tabular}
\end{table}
  We list the inner and outer disc radii and total disc mass for
each model (designated E,N,P1 and P2) disc in Table 1.
The extended (E) and narrow (N) simulations share the same inner (`cavity')
radius but differ in their outer radii; the mean emissivity per unit mass
is higher in the narrow model (on account of its higher  mean temperature) and thus the total
disc mass required for a fixed mm flux is somewhat lower. 
In addition,  we  compute a couple of variant prescriptions for the mm
emission, motivated by the results of recent simulations  by Pinilla et al (2012).
These relax the assumption of constant gas to dust ratio and follow the
evolution of the grain size distribution and spatial variation of the
dust in the case of a disc whose gas density profile is sculpted
by a planet. Dust is concentrated in the resulting structure 
within 
a pressure bump
located at about twice the orbital radius of the planet, with 
the disc being  strongly depleted in dust at radii interior to  this
pressure bump (we however note that
these
dust calculations are run for a small fraction of a viscous time
and thus - since the disc has not evolved into a steady state -
the results should be regarded as somewhat provisional). We model
this situation by two crude approximations that are intended to
bracket the simulation results. In model P1, the planet orbital
radius is halved with respect to the default model (i.e. $25$ A.U.
compared with $50$ A.U.) and the region between $50$ and $25$ A.U.
is filled with dust-free gas with a  surface
density profile that is an extrapolation of the power law profile; 
this increases the total gas mass
by a factor two  with respect to the  model N (which shares the same
outer radius). The gas
hydrodynamics and planetary orbital migration is modelled exactly as before;
as the planet migrates, it is assumed that the inner edge of the
dusty disc remains at
twice the instantaneous planet orbital radius.
In model P2, the disc gas is again extrapolated to the planet location
($25$ A.U.); emission is again only calculated from outward of the
cavity radius (i.e. twice the
instantaneous radius of the planet) and the initial outer radius is again
$75$ A.U.. In this case, however, the flux
contribution from 
dust that would have been  located between the planetary radius and twice this
radius 
is calculated as optically thin emission at the temperature of the
cavity radius. The placing of additional emission at the cavity
radius increases the mean emissivity per unit gas mass compared with model
N and consequently model P2 has a modestly lower total mass in order to
reproduce the same mm flux.

  In summary, each of these models are designed to reproduce observational
parameters (cavity size of 50 A.U. and mm flux of $\sim 100$ mJy) that
are typical of large hole (mm bright) transition discs.
We then evolve the coupled disc-planet system for a range of
different planet masses, 
and track the evolution of the system in the plane
of mm flux versus cavity radius.
We emphasise that at this stage we generically describe the companions 
as `planets' even though we shall include companions with masses up to
$100$ Jupiter masses. We do not extend our calculations to higher masses
on the grounds that Type II migration theory (which places the centre of
mass of the system at the primary star) becomes inapplicable at higher
mass ratios. Thus we cannot directly address the disparity in mm fluxes between
large hole transition discs and stellar binaries.

\section{Results}

  As the planet and disc inner edge  migrate inwards, the mm
flux changes due to three effects: redistribution of material in radius (and hence
temperature), optical depth effects and mass loss from the outer disc
due to the leakage flow to the inner disc (which
is assumed not to contribute to the mm flux). The two latter
effects both result in a reduction in mm flux. The former
can change the mm flux in either direction since viscous evolution
results in material spreading both inwards and out - i.e. into both hotter and cooler regions. 
In practice, we find that the net effect is either rough constancy of
the mm flux or else a gentle fading as the planet migrates inwards.
We find that these two outcomes depend on the relative masses of the planet
and the disc. In the case of a planet that is comparable to or less
massive than the disc, the planet is conveyed inwards as though it were
a representative fluid element in the disc; the disc structure upstream of
the planet is not significantly modified by the planet's presence and the
mm flux is nearly constant as the planet and associated disc hole moves
inwards. This behaviour is seen in models 100 P1, 40N and 10 N in Figure 1 
(where the number refers to the planet mass - in Jupiter masses - and
the model designation is defined in Table 1). On the other hand, in models
where the `planet' is more massive than the disc (such as 100 E, 100 N
and 100 P2 in Figure 1), the behaviour is somewhat different since  the
finite inertia of the planet impedes the free viscous migration
of the disc inner edge (Lin \& Papaloizou 1986, Syer \& Clarke 1995, Ivanov, Papaloizou \& Polnarev
1999). The slower migration means that there
is time for a significant depletion of the outer disc by the leakage
flow and  more than half the initial disc mass has leaked
past the planet in these models by the time it
reaches $10$ A.U..  The mm flux declines by more than a factor two
over this time, with additional fading resulting from the disc's
expansion to large radii
where the temperature - and associated mm emission - is low.

 Figure 1 plots observed transition discs in the plane of mm flux  
(scaled to a distance of $140$pc) versus hole size (see Owen \& Clarke
2012 for details of the mm data which is mainly obtained from the
mm surveys of Andrews \& Williams 2005,2007, Henning et al 1993 and
Nuernberger et al 1997). In the minority of
objects that lack $1.3$ mm fluxes, this is converted from $800 \mu$m data
using the prescription of Cieza et al 2008. 
The open circles denote 
systems that have been imaged by Brown et al. (2009) \& Andrews et al (2011) and which 
therefore represent the systems with the largest holes and highest mm fluxes. 
For the remaining unresolved objects the hole radius is either obtained from 
detailed SED modelling (Andrews et al. 2012, Calvet et al 2002,2005, Espaillat et al 2007,2010, 
Kim et al 2009, Merin et al 2010, Najita et al 2007: shown as open symbols) or, in the case of filled symbols,  is simply estimated from the `turn-off 
wavelength' listed in Cieza et al (2010); the hole
radius is thus more uncertain in these latter systems. The squares and triangles distinguish mm detections from upper limits.

\begin{figure}
\centering
\includegraphics[width=\columnwidth]{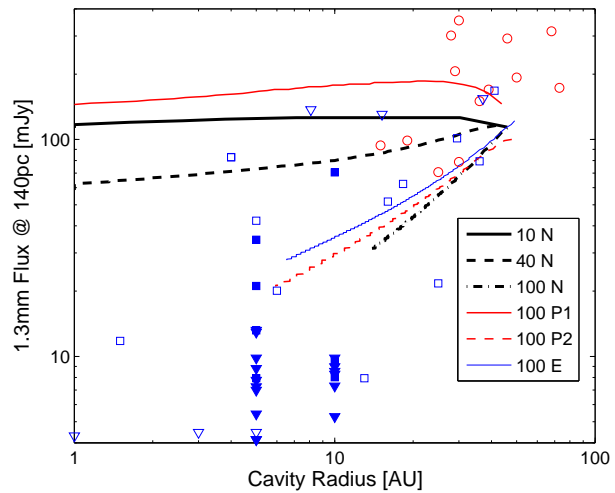}
\caption{Evolution in the plane of hole radius versus mm flux
for a variety of models labelled in each case with the `planet' mass (in
Jupiter masses)  and disc model as detailed in the Table. The  datapoints
denote observed transition discs (symbols detailed in the text). Millimetre
fluxes are re-scaled to the distance of Taurus using Spitzer c2d distances to
each region.}
\label{hereifany3}
\end{figure}

 Figure 1 illustrates that
there is a lack  
of mm bright objects (with flux
$> 30$ mJy at the distance of Taurus) with hole sizes $< 10$ A.U..
Although such objects are obviously not the targets of mm imaging studies,
they would have been readily picked up in photometric mm surveys and
there are likewise no reasons why such objects would not be identifiable
as transition discs from their SEDs (see Owen \& Clarke 2012).
This observational constraint defines the range of models that provide an
acceptable fit to observations.
Evidently it is only models with a rather large planet to disc mass ratio (a factor
two or more) that avoid evolving into the forbidden region with high mm flux
and small hole size. Furthermore, we emphasise those models that remain mm bright at $10$ A.U. spend a comparable time with hole sizes in the range $50-10$ A.U. and $10-1$ A.U.. We thus
cannot appeal to rapid inward migration at $< 10$ A.U.
in order to explain the observed lack of objects in the forbidden zone.
Given the requirement that the disc has to be massive enough
to generate the observed mm fluxes of large hole, mm bright systems ($\sim
100$ mJy) this in practice rules out systems in which the `planet' is less
than $\sim 100$ Jupiter masses (i.e. it excludes all companions in the planetary mass
or brown dwarf regime).

\section{Discussion}
 
 Our results above imply that  the model in which large mm bright
transition discs are associated with a `planetary' companion
is viable only if the companion is in fact of stellar mass. 
This is simply because less massive companions are
swept to small radii while the system remains mm bright, thus
contradicting the observational dearth of small, mm bright holes.
Our initial conditions are informed by the observed high mm fluxes of  
large cavity
transition discs 
so that  one cannot avoid this
conclusion by simply invoking lower mass outer discs.

 We noted above that we do not model the mass ratio regime of most stellar
binary companions. However, our results for a $100$ Jupiter mass
companion do not allow us to  explain  the observed  low mm fluxes in young stellar
binaries (Kraus et al 2011,2012) since they do not show a strong decline
of mm flux at large hole radius. This suggests that the low observed mm flux in
stellar binaries may more relate to the consumption of the disc
when the binary companion is formed rather than to the evolutionary
effects explored here. 
 
 We find that companions of around $100$ Jupiter masses provide a
good fit to the observed distribution of transition discs in the
plane of mm flux versus cavity radius, since such systems fade to less
than $30$ mJy by the stage that the hole size is $\sim 10$ A.U.. We
are not concerned that such systems would not fade to the  lowest
mm flux levels among  transition discs with small inner holes since a
separate mechanism - e.g. photoevaporation - could be invoked to explain
the faintest objects. 

 Nevertheless, there is an unassailable objection to
invoking  companions of around $100$ Jupiter mass: such objects would be
readily detected by imaging surveys (whose current sensitivity levels
extend to objects of $\sim 10$ Jupiter masses or lower (Kraus et al 2011,
2012). The absence of such companions in transition discs,
combined with the requirement demonstrated here of a rather massive
companion, is a serious challenge to the notion that transition discs
are associated with companions in {\it any} mass range. (See also Zhu et al
2011,2012 for other arguments against the planetary hypothesis for
the origin of transition discs based on difficulties in reproducing
the spectral energy distribution).   

\section{Conclusion}

  We  conclude that the popular planet model for  large cavity 
transition discs is faced with a `planet mobility problem'. If we set
up a system with an outer disc mass that reproduces the mm flux
of large cavity transition discs and set a companion within the cavity,
then the planet should migrate inwards by Type II migration and the
cavity radius thus shrinks with time. We however find that both  
planets and brown dwarfs  (i.e. objects less than $\sim 100$ Jupiter masses) 
are swept
to small radii by Type II migration 
{\footnote {Note that this conclusion is qualitatively
compatible with the findings of Armitage \& Bonnell 2002}}
and that the mm flux of the disc does
{\it not} fade significantly during this process. Thus we would expect to
see an associated population of mm bright objects with small holes 
($< 10$ A.U.) which
are {\it not} observed. We can avoid this outcome by instead invoking
a more massive companion (i.e. a low mass star). In this case the
migration is slow enough for the disc to fade at mm wavelengths
before the hole shrinks to 10 A.U.. However, such massive companions
in transition discs are clearly ruled out by recent imaging surveys
(Kraus et al 2011,2012).
 
 In order to `rescue' the planet scenario, we need some mechanism
that stops the planet migrating inwards and/or suppresses
the production of mm flux as the planet migrates. Photoevaporation
might appear to be an attractive scenario in both respects
(Rosotti et al 2013); however the initial conditions that are required
to match the high mm fluxes of large cavity transition discs
imply massive discs, so that the photoevaporation time-scale
would be long ($ \sim$ a Myr) even for systems with the highest
X-ray luminosity.

  Perhaps a more likely explanation is that there is still much to learn about
the secular evolution of coupled planet/disc systems. This issue 
is particularly acute because the low mass planets that would be
compatible with the null results from imaging surveys are in the regime
where the leakage flow could play an important role in slowing
planetary migration. On the other hand, it is not clear whether
dust filtration  - as is necessary to produce a transition disc
signature - would be effective in the limit that the flow past
the planet is almost unimpeded. These are issues which
can only be assessed by
future 2D/3D simulations exploring the secular evolution of
coupled planet/disc systems.

\section{Acknowledgments} We are grateful to the referee, Richard
Alexander, for an insightful report which has helped us improve the paper.

{}
\end{document}